\documentclass[]{aastex631}

\begin{document}

\title{Impact of High Intensity Long-Duration Continuous Auroral Electrojet Activity (HILDCAAs) on relativistic electrons of the Radiation Belt of Earth during Van Allen probe era}

\author[0009-0009-8688-7135]{Ayushi Nema}
\affiliation{Sardar Vallabhbhai National Institute of Technology \\
Surat, India}

\author[0000-0003-4281-1744]{Ankush Bhaskar}
\affiliation{Space Physics Laboratory, ISRO/Vikram Sarabhai Space Centre, Thiruvananthapuram, India}

\author[0000-0001-9365-6137]{Kamlesh N. Pathak}
\affiliation{Sardar Vallabhbhai National Institute of Technology \\
Surat, India}

\author[0000-0002-0116-829X]{Smitha V. Thampi}
\affiliation{Space Physics Laboratory, ISRO/Vikram Sarabhai Space Centre, Thiruvananthapuram, India}

\author[0000-0002-5333-1095]{Abhirup Datta}
\affiliation{Department of Astronomy, Astrophysics and Space Engineering, Indian Institute of Technology Indore, India}

\begin{abstract}

This study investigates the impact of High-intensity Long-Duration Continuous Auroral Electrojet Activity (HILDCAA) on the relativistic electrons in radiation belt of Earth. Utilizing data from Van Allen Probe mission of NASA, we conducted a comprehensive statistical analysis to understand the impact of HILDCAA events on the radiation belt fluxes. The super epoch analysis was carried out to determine the general response of L-shell, pitch angle, and energy-dependency of relativistic electrons to HILDCAAs. The analysis reveals a significant flux enhancement in the relativistic electron fluxes, predominantly occurring with a delay of 0 to 2 days following the onset of HILDCAA events. The general response indicates that the maximum energy of accelerated electrons reaches up to 6 MeV. Additionally, electrons with perpendicular pitch angles exhibit a significantly greater enhancement in flux and achieve higher maximum acceleration energies compared to those with parallel pitch angles. The observed time-delayed and pitch angle-dependent response related to the onset of HILDCAAs highlights the significant influence of wave-particle interactions, particularly in relation to ultra-low frequency (ULF) waves in this context. This is further supported by ground-based magnetometers and in-situ magnetic field observations from the RBSP probe, which demonstrated enhanced power of ULF waves during HILDCAA events. The study strengthens our current understanding of radiation belt particle acceleration processes and has potential implications for satellite operations and other space-based technologies, both on Earth and in the magnetospheres of other planets.

\end{abstract}


\section{Introduction} \label{sec:intro}

The acceleration of electrons in the outer radiation belt of the Earth remains a complex and intriguing topic for researchers. The existence of this belt of charged particles was first confirmed by the Explorer I mission, leading to its subsequent naming in honor of James Van Allen. The Van Allen outer radiation belt primarily consists of particles originating from the solar wind that become trapped by the planet's magnetosphere. The Van Allen Probe mission, also known as Radiation Belt Storm Probes (RBSP)  \citep{breneman2022van} launched by NASA aimed to study the behavior of these trapped particles in response to changing solar activity and transient events. The outer radiation belt is highly dynamic and varies minute to solar cycle time scales \citep{baker2014relativistic, kozyreva2009variations,hajra2014relativistic}. The solar wind transients like coronal mass ejections, Corotating Interaction regions, shocks, etc, \citep{ murphy2018global,pandya2019variation, schiller2018cause,bhaskar2021radiation} and geomagnetic activity within magnetosphere like substorm, plasma waves are primary causes of the modulation of the particle fluxes within the radiation belts \citep{ma2022effects, summers2004modeling,fok2001rapid}.  \cite{freeman1964morphology} observed that high-energy (MeV) electron fluxes dropped during the storm's main phase and increased back to or above pre-storm levels during the storm's recovery phase over a significant portion of the outer radiation belt (L$>$4). Also, another study by \cite{brautigam2000radial} suggests that radial diffusion \citep{osmane2021radial,ozeke2014analytic, ozeke2012ulf} of relativistic electrons propagates outer boundary variations into the heart of the outer radiation belt during the storm recovery phase, leading to both significant decreases and increases in the $<$1 MeV electron fluxes throughout that region. However, the electron fluxes could drop rapidly by several orders of magnitude and stay low even after the storm is over \citep{turner2012explaining}.

 Even without storms, enhanced whistler-mode chorus waves during prolonged substorm activity can generate relativistic electrons in the Earth's outer radiation belt in general \citep{ma2022effects, murphy2018global}. \cite{mathie2000correlation} reported a rare radiation belt occurrence in which there were no discernible VLF chorus waves but strong ULF waves. An average augmentation of relativistic electron fluxes was observed, which, within the context of the VLF chorus wave-driven local acceleration, might be readily misunderstood. The thorough simulation and the high-resolution data demonstrate that the ULF waves could radially spread and accelerate the radiation belt electrons effectively in the absence of VLF chorus waves. Even \cite{mann2016explaining}'s study showed how the third radiation belt formed as a direct result of the extraordinarily rapid outward ULF wave transfer during storms. According to \cite{fukizawa2020pitch}, there is a statistically significant correlation, indicating that the dynamics of electron precipitation into the atmosphere are influenced by electron cyclotron harmonic waves. This new viewpoint implies that the dynamics of the ultra-relativistic third radiation belt may not require high-frequency wave-particle scattering loss into the atmosphere \citep{murphy2020inner}.

There exists a distinct class of solar wind structures that significantly impacts the Earth's magnetosphere-ionosphere system over extended periods. These structures are characterized by prolonged intervals of fluctuating southward components of the interplanetary magnetic field (IMF \( B_z \)). When the southward \( B_z \) component persists, it facilitates the transfer of solar wind energy into the magnetosphere, resulting in sustained auroral activity. This phenomenon is referred to as High-Intensity Long-Duration Continuous Auroral Electrojet Activity (HILDCAA) \citep{tsurutani1987cause, tsurutani2004high, soraas2004evidence, guarnieri2006nature, hajra2013solar, hajra2014superposed}.  These events play a crucial role in geomagnetic activity in the auroral region, occurring without significant development of ring current. HILDCAA events are geomagnetic disturbances characterized by intense auroral activity and thus sustained high levels of the AE/AL index, often exceeding 1000 nT amplitude. These events typically last for a minimum of two days, during which the AE index remains above 200 nT without dropping below this threshold for more than two hours. Unlike geomagnetic storms, HILDCAAs occur outside of the main storm phases and are primarily known for their prolonged and continuous nature. 

HILDCAA events are particularly noteworthy for their impact on the Earth's radiation belts, especially on relativistic electrons within the energy range of approximately 1 to 5 MeV. The continuous substorm activity associated with HILDCAAs generates plasma waves such as ULF (ultra-low-frequency) waves, which play a critical role in the acceleration and transport of these high-energy electrons. The frequent and intense auroral activity during HILDCAAs is a key driver of these ULF waves, leading to significant changes in the dynamics of the radiation belts\citep{hajra2015relativistic_CIR, su2015ultra}. As mentioned by \cite{summers2004modeling} HILDCAA events during which prolonged substorms and enhanced chorus activity occur, are associated with relativistic electron flux enhancements. At geosynchronous orbit, the HILDCAA events are found to be related to an enhancement of relativistic (E $>$ 2 MeV) electron fluxes in the magnetosphere. Recently, \cite{hajra2024ultra} studied ultra-relativistic electrons of energy $\sim$2-20 MeV. They analyzed electron acceleration for HILDCAA events based on their duration. HILDCAA events of longer periods are associated with larger energy of accelerated relativistic electrons. At the onset of HILDCAA occurred in 2019 February 1-3, relativistic electron fluxes of magnitude $\sim$2.0-2.9 MeV declined from the magnetosphere at around L$>=$4. The study indicated that electron depletion is possibly due to the magnetopause shadowing effect and electron interaction with EMIC waves. According to \cite{milan2023solar}, every HILDCAA onset is accompanied by substorm-specific changes in open magnetic flux, reconnection rates on both the dayside and nightside, the cross-polar cap potential, and the AL index. Their study highlights the close relationship between substorm activity and AE/AL disturbances during HILDCAAs, suggesting that the magnetic spikes observed during these events are directly associated with jumps in the AE/AL indices and the generation of ULF waves.

Thus, based on past studies, HILDCAA events are strongly correlated with increases in relativistic electron fluxes in the outer radiation belt\citep{hajra2015relativistic}. However, despite the growing understanding of HILDCAA events, several questions remain open, particularly regarding their general and long-term impact on radiation belt relativistic electrons. How do these events influence the dynamics of relativistic electrons over extended periods? What role do ULF waves play in the acceleration and loss of these particles during HILDCAA events? Addressing these questions is crucial for understanding the broader implications of HILDCAA events on space weather and the Earth's radiation environment.

The present study attempts to address some of these open questions by investigating the impact of HILDCAA events on relativistic electrons in the Earth’s radiation belt specifically during NASA's Van Allen Probes era. By analyzing in situ particle measurements for all HILDCAAs occurring in this period, we seek to shed light on the mechanisms through which they influence the behavior of high-energy particles, contributing to a better understanding of the complex interactions between geomagnetic activity induced by HILDCAAs and radiation belt dynamics.

\section{Data} \label{sec:Data}
The HILDCAA events have been identified by following the strict criteria given by Tsurutani and Gonzalez \citep{tsurutani1987cause}. The AE and Sym-H indices have been obtained from \url{https://wdc.kugi.kyoto-u.ac.jp/}. The near-earth solar wind conditions during HILDCAA have been studied using the data available at \url{https://omniweb.gsfc.nasa.gov/}. The OMNIWeb database is already time-shifted to the bow shock's nose and the measured quantities in the GSM coordinate system have been used for further analysis \citep{russell1971geophysical, hapgood1992space}.

For the radiation belt particle measurements, primarily we have used the observations from the twin spacecraft Van Allen Probes, which are in proximity to the equatorial plane(\url{https://rbsp-ect.newmexicoconsortium.org/data_pub/rbspa/rept/level3/pitchangle/}). For the current analysis,  we have used only RBSP-A spacecraft data.
The relativistic electron fluxes in differential energy channels have been obtained from the Relativistic Electron Proton Telescope (REPT) instrument \citep{baker2014relativistic} aboard Van Allen Probes. With 12 energy channels and 17 pitch angles, the REPT instrument provides highly resolved electron observations in terms of energy and pitch angle. Since their launch in 2012, the Van Allen Probes have observed a total of 13 HILDCAA events, which are summarized in Table~\ref{tab:Table 1}. For this study, the relativistic electron flux data corresponding to these identified HILDCAA events has been sourced from the Van Allen Probe Science Gateway, available at \url{https://rbspgway.jhuapl.edu/}.

\begin{table}[]
    \centering
    \begin{tabular}{c c c c c c c c}
    \hline
    \hline
     Event number & Start date & Start time & End date   & End time &  Peak V\textsubscript{sw} & Peak AL   & Peak SymH \\
                  &            &   (UT)     &            &   (UT)   &         (km/s)            &    (nT)   &   (nT)  \\
        \hline
                1 & 2014/09/23 &  06:40     & 2014/09/25 & 14:15    & 574.292                   & -784.72   & -32.0           \\
                2 & 2015/10/01 & 12:09      & 2015/10/03 & 19:57    & 779.864                   & -1143.408 & -120.667        \\
                3 & 2015/12/10 & 00:13      & 2015/12/12 & 00:28    & 739.077                   & -1109.912 & -110.170        \\
                4 & 2016/10/16 & 08:12      & 2016/10/18 & 11:05    & 667.8                     & -1031.0   & -50.023         \\
                5 & 2016/11/24 & 00:00      & 2016/11/26 & 12:03    & 748.563                   & -1261.918 & -50.435         \\
                6 & 2016/12/21 & 09:41      & 2016/12/24 & 18:57    & 756.424                   & -733.667  & -41.914         \\
                7 & 2017/01/31 & 08:34      & 2017/02/04 & 01:01    & 692.806                   & -652.944  & -39.069         \\
                8 & 2017/07/24 & 03:26      & 2017/07/26 & 19:06    & 753.843                   & -926.26   & -48.64          \\
                9 & 2017/08/17 & 07:28      & 2017/08/21 & 10:24    & 670.1                     & -752.054  & -59.061         \\
                10& 2018/05/08 & 14:44	    & 2018/05/11 & 12:43    & 708.293                   & -705.0    & -41.109         \\
                11& 2018/05/31 & 14:36      & 2018/06/03 & 20:28    & 622.325                   & -699.581  & -33.698         \\
                12& 2019/02/01 & 02:38	    & 2019/02/03 & 09:48    & 594.234                   & -726.625  & -42.106         \\
                13& 2019/02/28 & 05:50      & 2019/03/02 & 21:25    & 632.4                     & -1063     & -45.0           \\
         \hline
         \hline
             \end{tabular}
    \caption{HILDCAA events list from the year 2014 to 2019 with peak solar wind geomagnetic parameters during the occurrence period.}
    \label{tab:Table 1}
\end{table}

For the space-based ULF wave analysis during HILDCAA events, we utilized the level-2 data from the Electric and Magnetic Field Instrument Suite and Integrated Science (EMFISIS), accessible through the Van Allen Probe Science Gateway at \url{https://emfisis.physics.uiowa.edu/Flight/RBSP-A/L2/}. Additionally, for the ground-based ULF wave analysis, we employed data from the IMAGE network of ground magnetometers, which can be found at \url{https://space.fmi.fi/image/www/index.php?page=user_defined}.

\section{Observations}\label{sec:Results}
\begin{figure}[ht]
    \centering
    \includegraphics[width=6.5in, height = 8in]{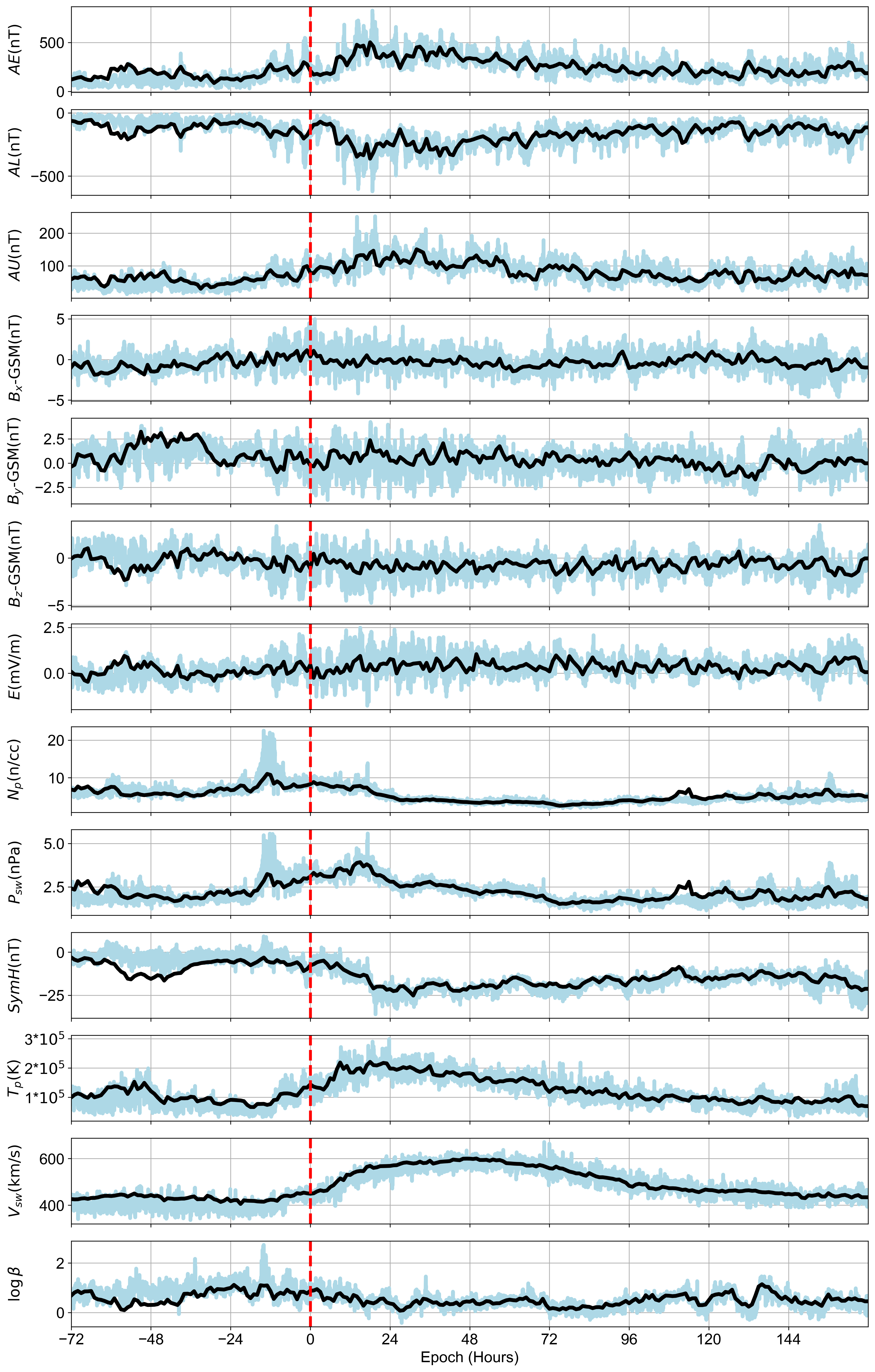}
    \caption{Average variation of interplanetary and geomagnetic parameters for 13 HILDCAA events under 1-min (blue line) and 1-hour (solid black line) time resolution. The red vertical dashed line at zero epoch represents the onset of all HILDCAA events.}
    \label{fig:1}
\end{figure}

As mentioned in the introduction, geomagnetic storms, substorms, and interplanetary (IP) shocks are crucial drivers of the acceleration and variability of relativistic electron fluxes in the outer radiation belt \citep{reeves2003acceleration, schiller2018cause, bhaskar2021radiation}. In this study, we focus on examining the impact of High-Intensity Long-Duration Continuous Auroral Electrojet Activity (HILDCAA) events on relativistic electron behavior. Figure~\ref{fig:1} displays the averaged data of interplanetary (IP) and geomagnetic parameters at both 1-hour and 1-minute resolutions for all 13 HILDCAA events. For each event, we considered a period extending 3 days before the start date and 7 days after, resulting in a consistent 10-day window for all events. Subsequently, we applied the superposed epoch analysis method to derive the averaged characteristics of the HILDCAA events. In this analysis, the onset of each HILDCAA event is indicated by the zero epoch.

This 1-hour resolution superposed epoched plot of the studied variables gives us a broader view of how IP and geomagnetic parameters varied during a typical HILDCAA event. For a closer look, we have also overplotted the 1-min resolution data in Fig.~\ref{fig:1}. 

The figure illustrates the response of key parameters during the onset of the HILDCAA event. In the upper panel, the AU and AL indices reflect significant changes: the AE index shows a sharp increase while the AL index decreases, indicating intensified auroral currents. The presence of eastward electrojets is evident from the rise in the AU index.  Before the event, the interplanetary magnetic field components (B\textsubscript{x}, B\textsubscript{y}, B\textsubscript{z}) in GSM coordinates remain relatively stable, nu shows strong fluctuations as compared to pre-onset time. Following the zero epoch, a noticeable disturbance in B\textsubscript{z} suggests that the coupling between the solar wind and the magnetosphere has intensified. Additionally, the strongly fluctuating solar wind motional electric field E indicates enhanced convection within the magnetosphere. After the onset of HILDCAA, solar wind speed V\textsubscript{sw} increases, which is associated with prolonged geomagnetic activity linked to corotating interaction regions (CIRs). Solar wind density N\textsubscript{p} and dynamic pressure P\textsubscript{sw} also rise prior to the onset, contributing to magnetospheric compression and increased energy input. In contrast, the Sym-H index, which tracks the symmetric ring current, displays a decrease; however, this decline is not as pronounced as that observed during the main phase of a geomagnetic storm, remaining close to -30 nT, typical of geomagnetically quiet periods. Finally, the steady increase in proton temperature T\textsubscript{p} after the zero epoch aligns with the presence of a sustained, high-energy solar wind stream.

\subsection{Energy  and Pitch angle dependence }
\begin{figure}[ht]
    \centering
        \includegraphics[width=6.6in, height = 7in]{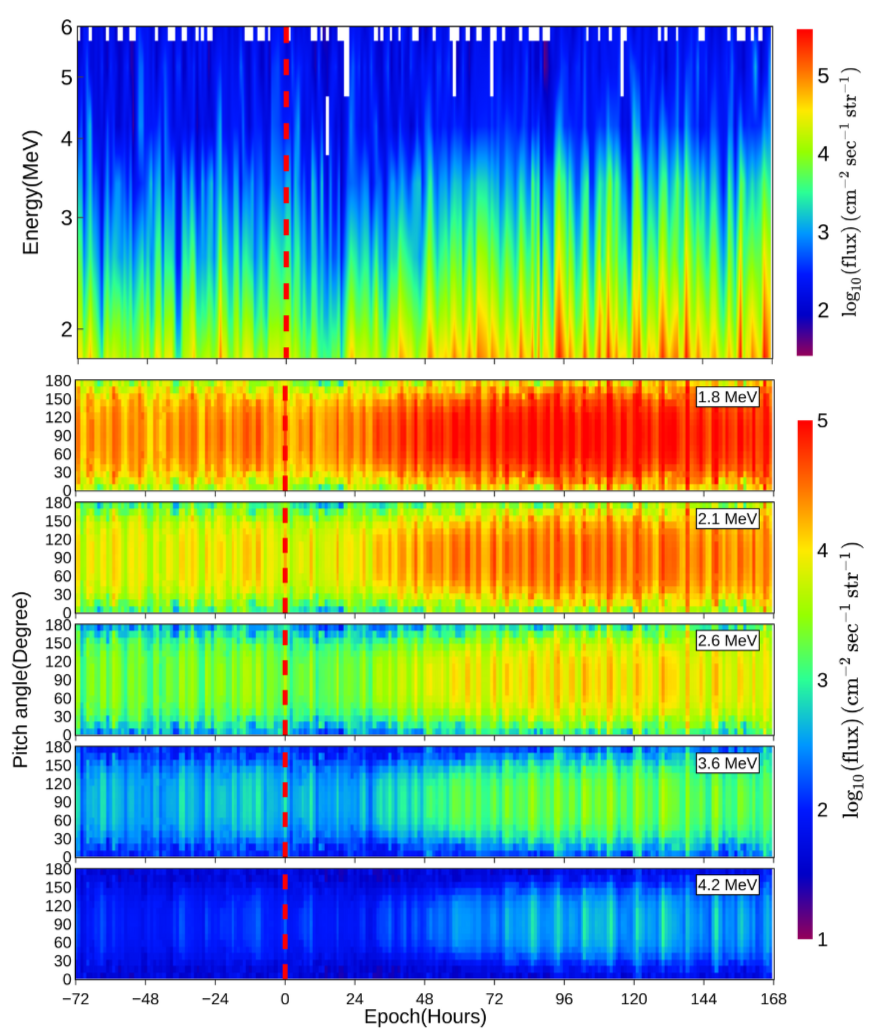}
    \caption{(Top panel) Superposed energy spectrum of e\textsuperscript{-} fluxes. (Bottom panel) Pitch angle variation of fluxes for different energies at L$>$4 shells. The vertical red-dashed line in the figure is the onset of all events.}
    \label{fig:2}
\end{figure}

In Fig.~\ref{fig:2}, the relativistic electron flux is presented using superposed epoch analysis of all HILDCAA events from 2014 to 2019. This analysis provides an average representation of the electron flux during these events, specifically for L$>$4 shell to indicate particle injection during substorms.  The top panel displays the Energy-Time spectrum of electron fluxes, with the red dashed vertical line marking the onset of the studied HILDCAA events. It is clear that electron fluxes increase significantly after a delay relative to the zero epoch, with enhancements varying by order of magnitude across different energy channels. Notably, this increase occurs approximately two days after the onset of the events, followed by a gradual peak and subsequent decay. In the bottom four panels, fluxes for five differential energy channels are plotted against pitch angles. These panels reveal distinct enhancements in flux, particularly at pitch angles between $60^\circ$ and $120^\circ$. The patterns observed in the bottom panels closely align with the gradual flux increases seen in the top panel, underscoring the relationship between pitch angle distributions and overall flux enhancements following HILDCAA events.

\subsection{ L-shell and Energy dependence of electron fluxes}
\begin{figure}[ht]
    \centering
    \includegraphics[width=4in]{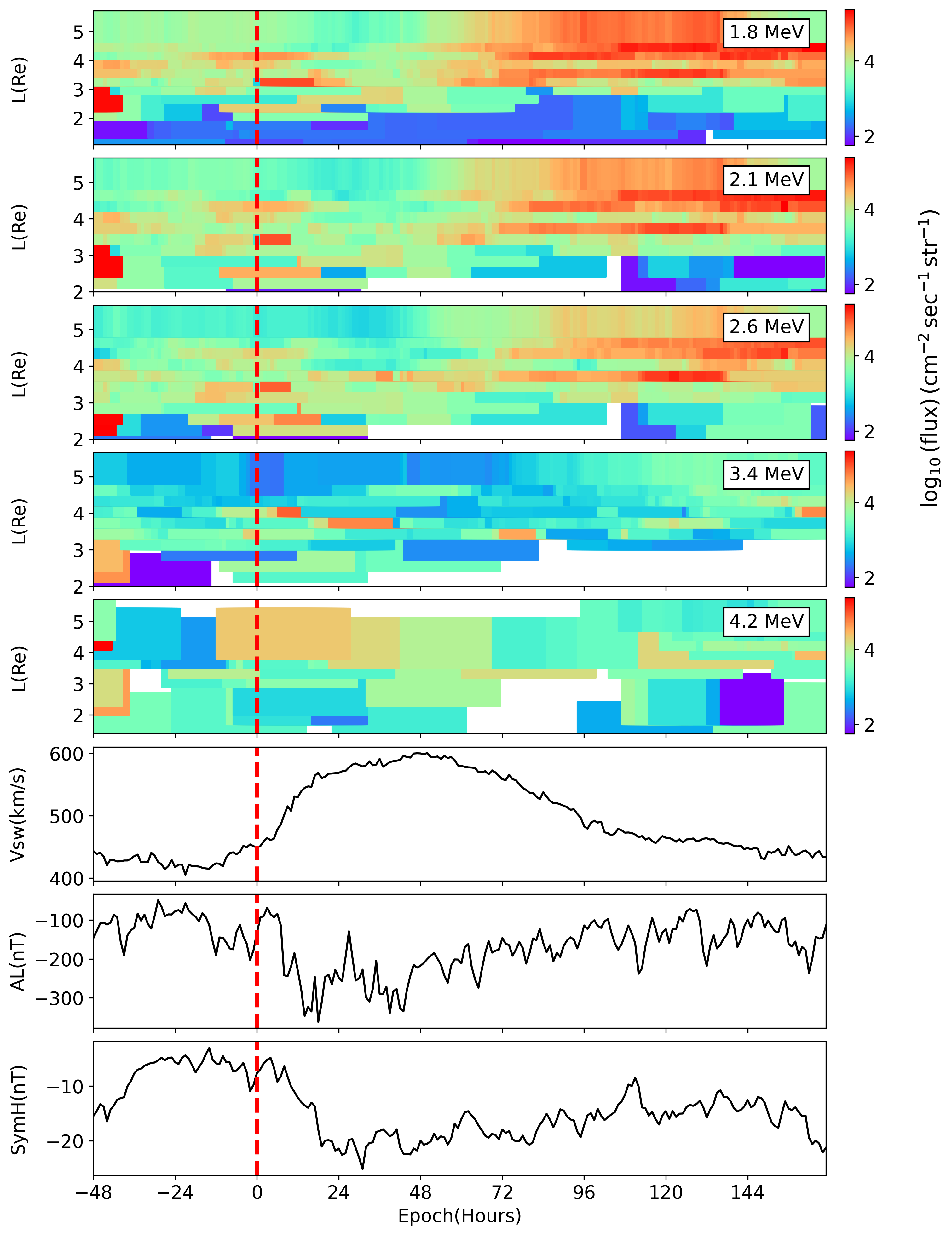}
    \caption{Superposed epoch analysis of electron fluxes w.r.t. L shells averaged over all pitch angles for five energy channels of REPT on Van Allen Probe-A. The electron fluxes show enhancement after a delay of 2 days. The red dashed line indicates the epoch time.}
    \label{fig:3}
\end{figure}

In Fig.~\ref{fig:3}, we present the results of a superposed epoch analysis of electron fluxes as a function of L-shell. Before this analysis, the fluxes were integrated over all pitch angles. To construct the Time-L-shell plot, we binned the data using a bin size of 0.3 L-shell and a 30-minute time bin to ensure adequate spatial and temporal resolution. Alongside the electron fluxes, we have also plotted the averaged solar wind speed, AL index, and Sym-H index to illustrate the relationship between solar wind dynamics, geomagnetic activity, and electron fluxes. The figure clearly shows that the electron fluxes in the post-epoch exhibit enhancements compared to the pre-zero-epoch period, with the 1.8 MeV energy channel displaying the highest fluxes at higher L-shells. Notably, the enhancements in electron flux were predominantly observed at higher L-shells. Furthermore, no significant changes were detected for energy channels above 4.2 MeV (higher energy channels not shown here). This observation raises two key questions: why does the enhancement in relativistic electron flux predominantly occur at higher L-shells, and why is this enhancement primarily observed in energy channels below 5 MeV? Additionally, we observed a delay of 2-3 days in the electron flux enhancement w.r.t peak solar wind speed (\(V_{sw}\)) and substorm activity, as indicated by the AL index (Fig.~\ref{fig:2}). To understand the cause of this delay, we investigated the role of ULF wave interactions during HILDCAA events, a topic that will be explored further in a subsequent section.

\subsection{L-shell dependence of field-aligned and trapped  electron fluxes}
\begin{figure}[ht]
            \includegraphics[width=3.5in, height = 6in]{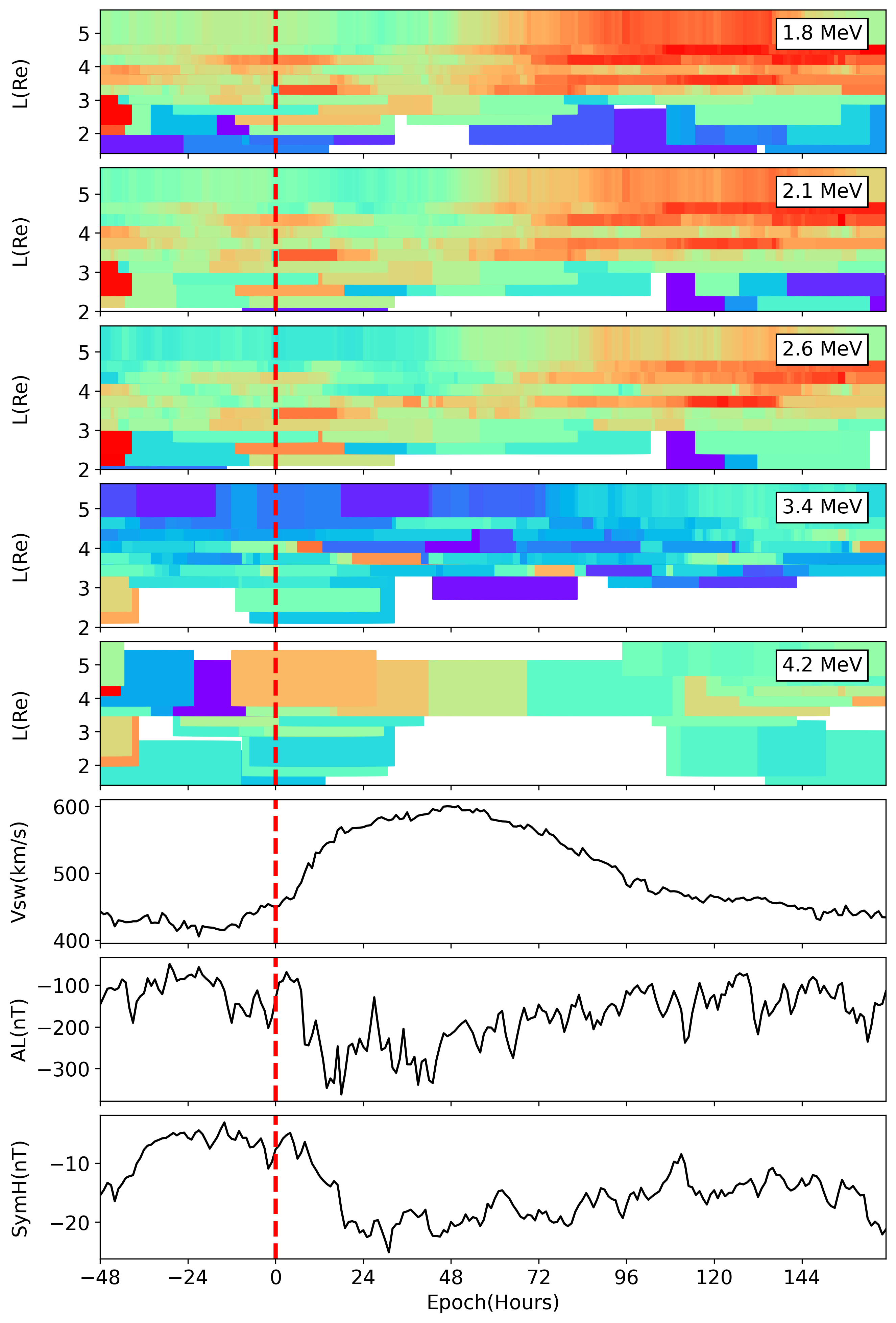}         
    \hfill
            \includegraphics[width=3.5in, height = 6in]{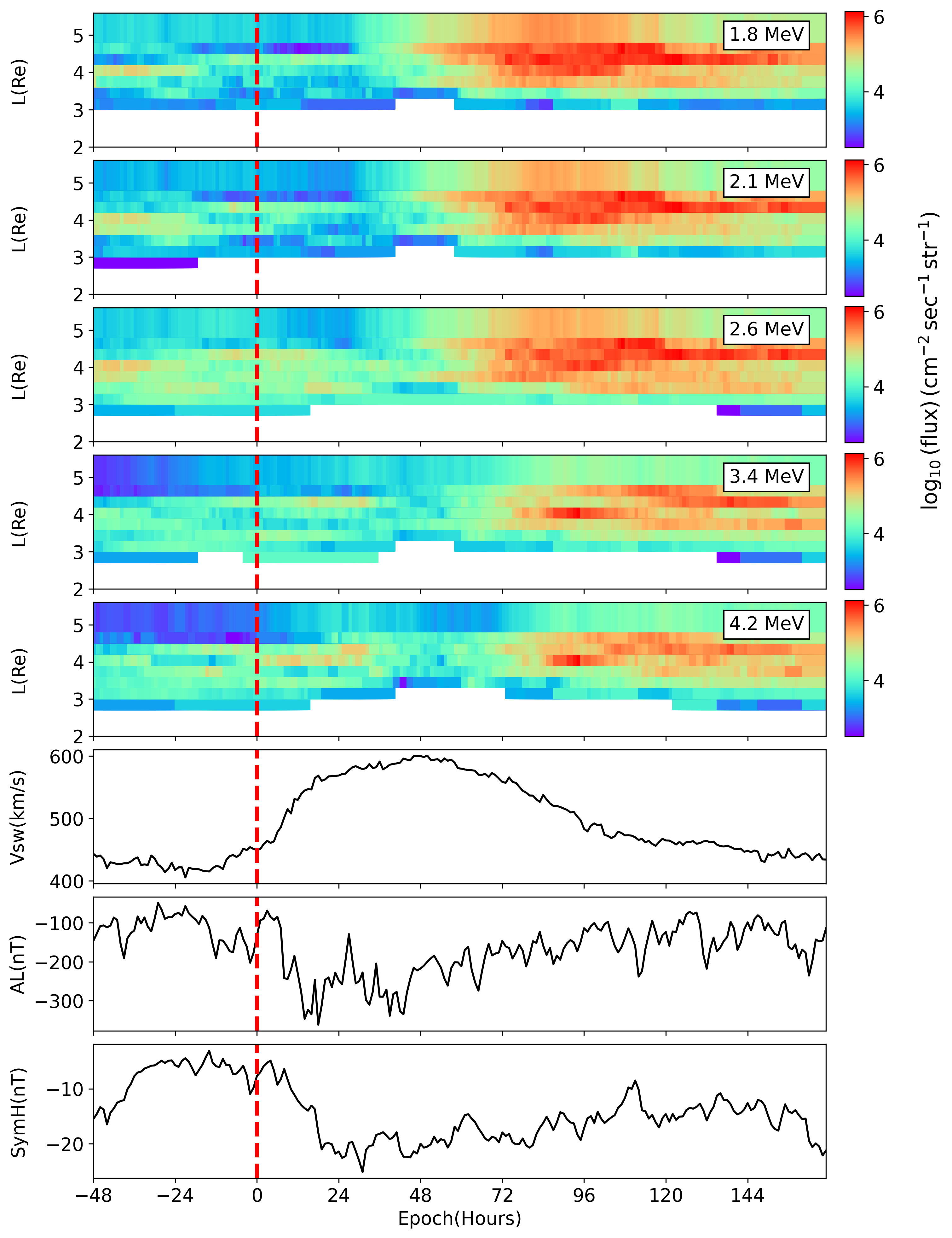}
            
    \caption{Flux enhancement between the nearly field-aligned (left) and nearly perpendicularly (right) trapped relativistic electrons in the outer radiation belt.}
    \label{fig:4}
\end{figure}

We conducted a superposed epoch analysis of field-aligned(5.29$^\circ$, 15.88$^\circ$ and 174.71$^\circ$) and nearly trapped(90.00$^\circ$ and 100.58$^\circ$) particles to statistically examine whether or not acceleration depends on the pitch-angle of electrons. In this analysis, we performed bin averaging of the flux data over each L-shell-Time bin. Each bin in the L-shell was set to a size of 0.3 L, spanning the range from 1 to 6 L, while the time bins were 30 minutes each. In Fig.~\ref{fig:4}, we compare near field-aligned and near perpendicularly trapped particles to investigate the potential pitch angle dependence of particle acceleration. 

Field-aligned particles have velocities that are predominantly parallel to the magnetic field lines, while perpendicularly aligned particles exhibit more complex motion as they travel toward or away from Earth, with their velocity components oriented perpendicular to the geomagnetic field lines.  In our analysis, perpendicularly aligned particles demonstrate greater energization compared to field-aligned particles. Specifically, the maximum energy observed for field-aligned particles reaches 2.6 MeV, whereas perpendicular particles can attain energies as high as 4.2 MeV, as shown in Fig.~\ref{fig:4}. Notably, there is a clear energy dependence of the delay in perpendicular particle fluxes compared to parallel fluxes. These findings are significant for understanding the underlying acceleration mechanisms, which will be further discussed in the in context of wave-particle interaction. 

Also, one can observe a shift in the delay of flux energization of electrons. The shift in the delay lasted longer for perpendicularly trapped electrons in comparison to field-aligned electrons.  

\subsection{ULF waves and their possible impact}
Ultra low-frequency waves (mHz), excited by the dynamic solar wind conditions, cause radial diffusion of particles\citep{claudepierre2008solar}. The rate of radial diffusion depends on the power spectral density of waves. It is known that ULF waves are generally responsible for the energy variation of the trapped particles. These waves are enhanced during the period of geomagnetically enhanced activity. Thus, it becomes necessary to study ULF waves during HILDCAAS to evaluate their impact on radial acceleration or loss of particles.

\begin{figure}[ht]
      \includegraphics[width=3.5in, height = 3in]{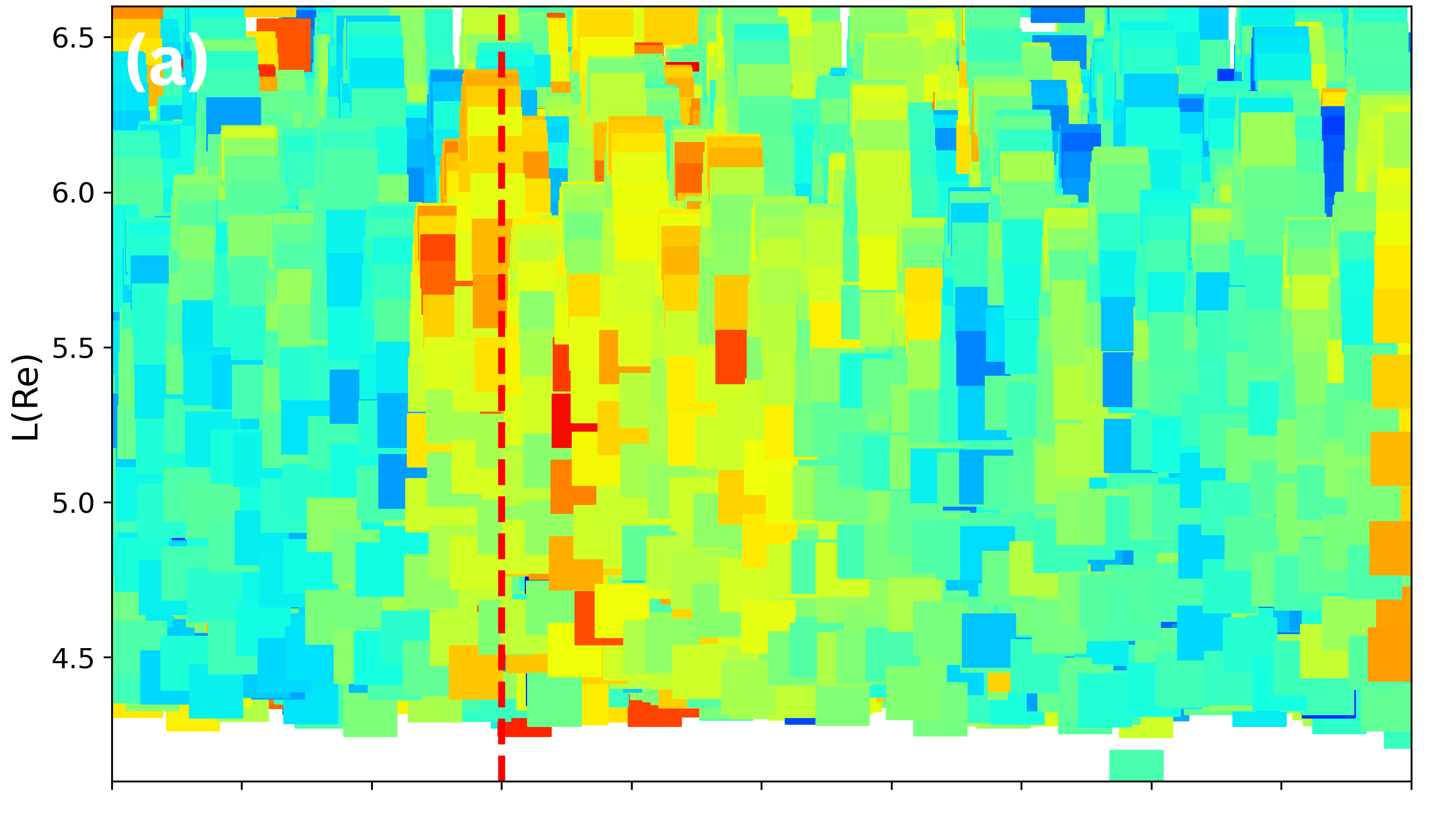}
    \hfill
      \includegraphics[width=4in, height = 3in]{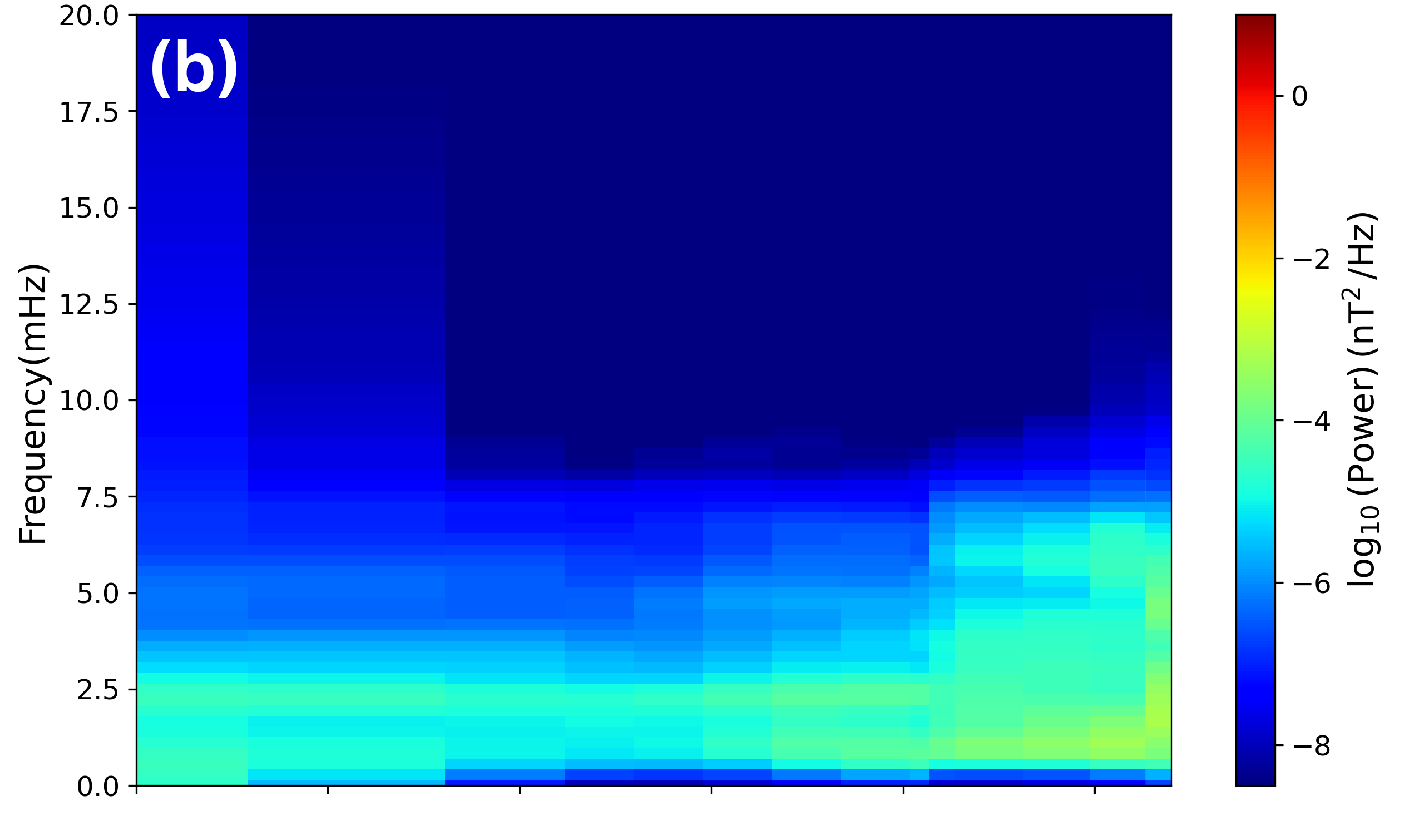}

  
      \includegraphics[width=3.5in, height = 3in]{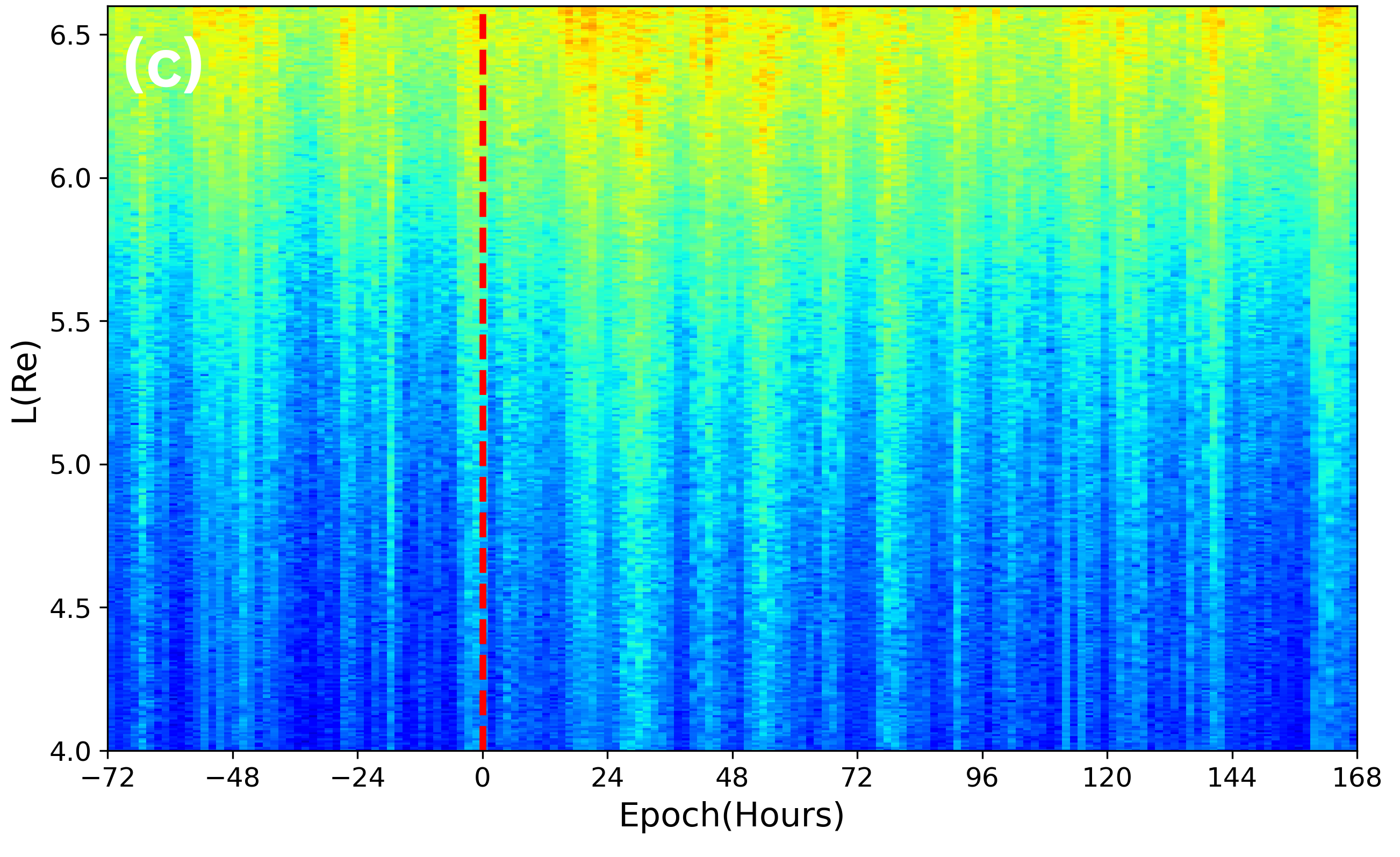}
   \hfill
      \includegraphics[width=4in, height = 3in]{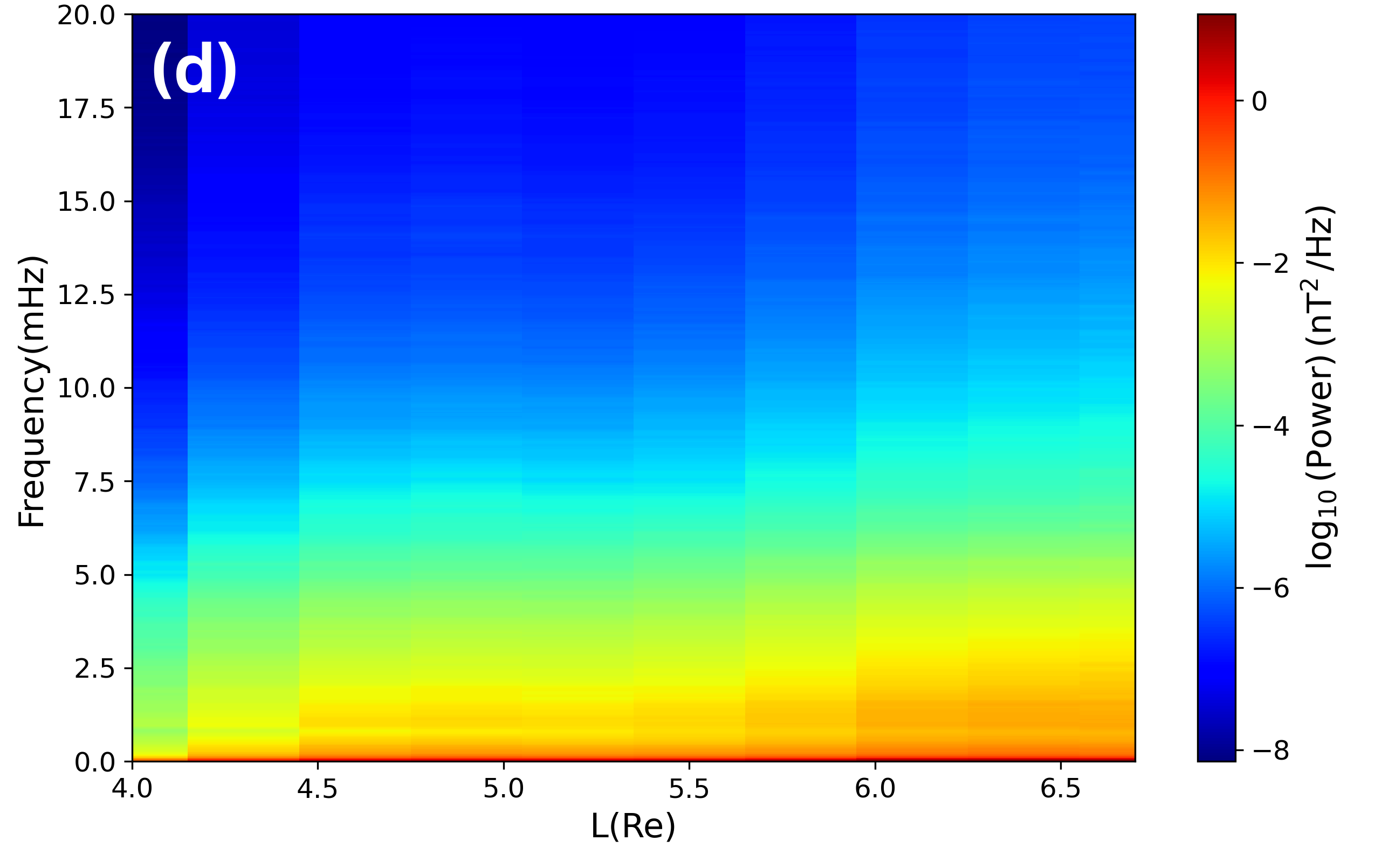}   
    \caption{ULF wave analysis from space and ground-based magnetometers. The two plots in the above panel are from EMFISIS and plots in the below panel are from the IMAGE magnetometer network. ULF power is observed to be enhanced during HILDCAA events.}
    \label{fig:5}
\end{figure}

 The magnetic field data from EMFISIS  was transformed into field-aligned coordinates. To unravel ULF waves,  the butter worth bandpass filtering having a band of 1-10 mHz was applied to this time series. The power was integrated over the ULF band for a given time. Further, the superposed epoch analysis was carried out for all studied HILDCAA events. In the top panels Fig.~\ref{fig:5} (a), the integrated power over 1-10 mHz is shown with respect to L-shells and time. It shows an enhanced power near the onset of events and during the initial activity of HILDCAAs. Fig.~\ref{fig:5} (b) shows the frequency response in the ULF range w.r.t L-shell. The presence of a few Hz waves at 4-6.8 L-shells, prominently at higher L-shells is evident. For higher L-shells, the power and frequency range seems to be higher.

The same pattern is followed by ground-based magnetometer data, provided by the IMAGE network. The power estimated from signals of ground-magnetomete station is under the same frequency range of 1-10 mHz as that of space based. Here also, the signals were filtered using butterworth band pass filter. The analysis is mainly done for L$>$4 shells. The red dashed vertical line differentiates how intensified power is before and after the onset of HILDCAA. Remarkably all space-based observations are consistent with ground-based observations as shown in (c) and (d). 


\subsection{Time lag correlation of electron flux and solar wind speed}
\begin{figure}[h]
\centering
        \includegraphics[width=3.5in, height = 3in]{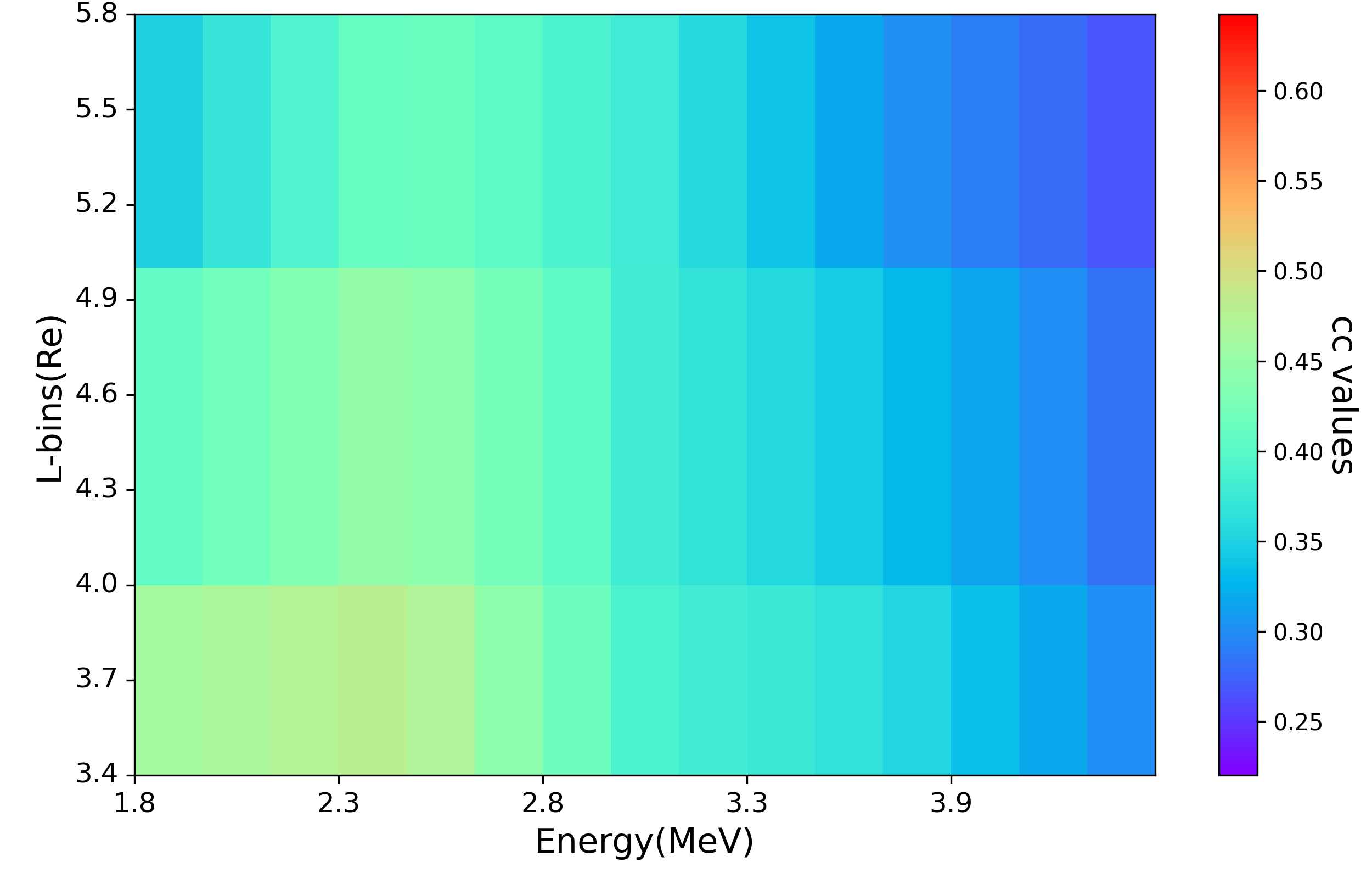}
        \hfill
        \includegraphics[width=3.5in, height = 3in]{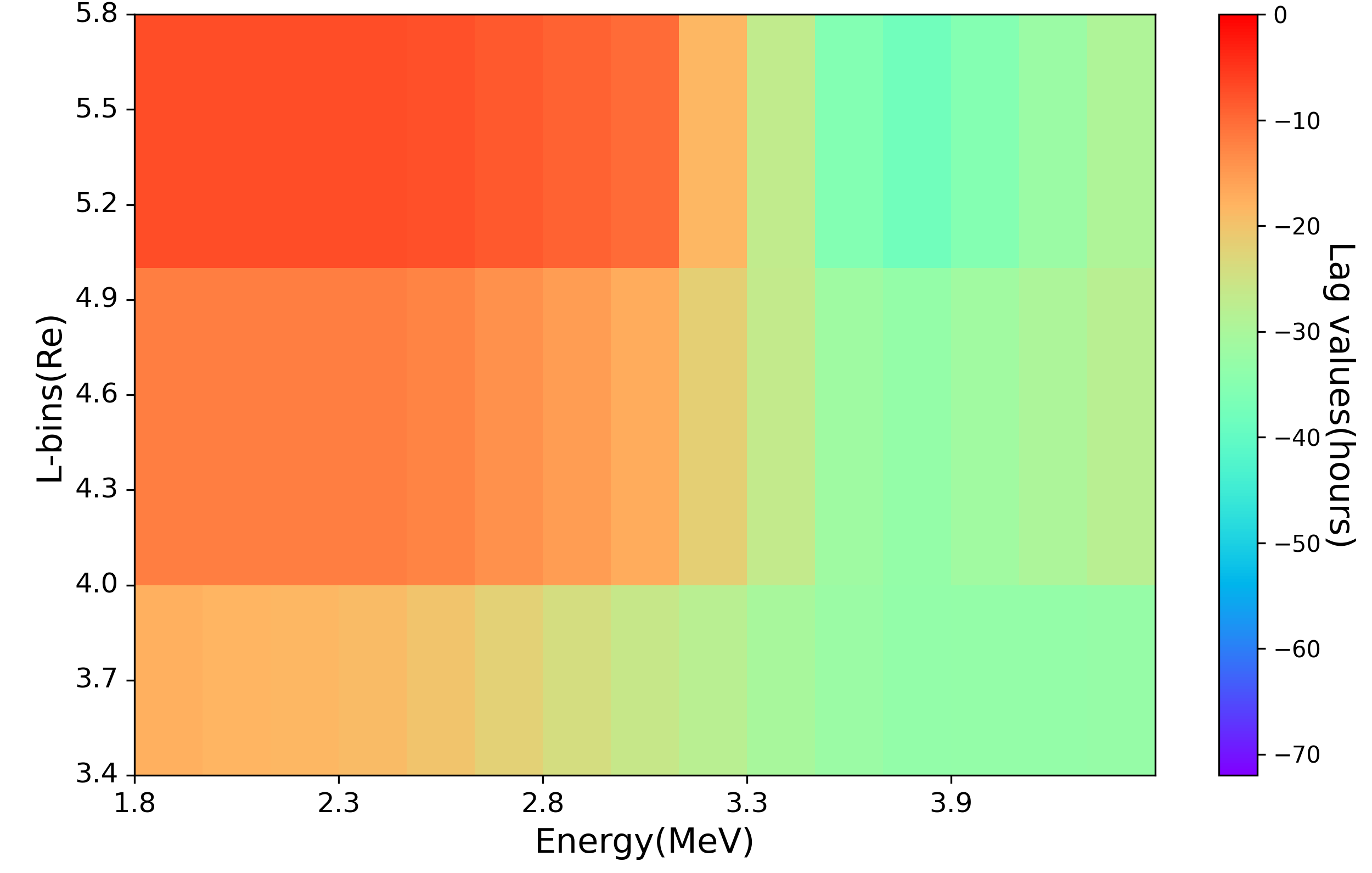}
    \caption{(Left)Pearson's correlation and (right)time lag derived between relativistic electron flux and solar wind speed.}
    \label{fig:6}
\end{figure}

Pearson's correlation and time lag derived between electron flux and solar wind speed has been shown in Fig.~\ref{fig:6} with respect to L-shell bins and energy respectively. The figure on the left shows the correlation of V\textsubscript{sw} and electron flux. Here we estimated Pearson's correlation between electron flux and solar wind speed for all 13 HILDCAA events for all energy channels from 1.8 to 4.2 MeV for L-shell bin size of 0.3 L. The analysis reveals that for lower energy channels at higher L-shells, there is a noticeable increase in correlation. Conversely, for higher energy channels at lower L-shells, the correlation between electron flux and solar wind speed decreases. Moreover, the figure the relationship between time lag and L-shells as well as energy channels. The right panel of Fig.~\ref{fig:6}, shows the time lag between electron flux and solar wind speed for L-shell and Energy grid. Notably, we observe that time lag is lower at higher L-shells and for higher energy channels. Conversely, the time lag increases for lower energy channels at lower L-shells. These observations highlight the complex relationship between these variables, suggesting that energy and spatial factors play significant roles in the acceleration of electrons during HILDCAAs.


\section{Discussion}\label{sec:Discussion}
In a study by \cite{anderson2015acceleration}, it was found that small storms can be just as effective as large storms in both enhancing and depleting electron fluxes in the Van Allen belts. The research revealed that the probability of flux enhancement increases with faster solar wind conditions, irrespective of the storm size. This study underscores the importance of examining cause-and-effect relationships in non-storm conditions to more accurately quantify electron acceleration and loss processes. This insight is particularly relevant when considering the impact of HILDCAA events, which, through prolonged and intense geomagnetic activity in high latitude, contribute to the observed enhancements in relativistic electron fluxes within the Van Allen belts though only weak ring current intensification present.

The present study investigated the impact of HILDCAAs on radiation belt electrons, specifically analyzing events that occurred during Van Allen Probes. The superposed epoch analysis shows delayed enhancements in relativistic electron fluxes, particularly at higher L-shells and in lower energy channels (below 5 MeV), with flux increases occurring 2-3 days after the event onset. Electron flux variations are observed to be pitch angle dependent, with greater energization for perpendicular particles than field-aligned ones.  Both space and ground-based data confirm enhanced ULF wave power near the onset of HILDCAAs, particularly at higher L-shells. Time-lag correlations between solar wind speed and electron flux reveal that higher energy channels at higher L-shells experience shorter delays, suggesting complex spatial and energy-dependent electron acceleration processes during HILDCAAs.

Particle flux enhancement in the magnetosphere is primarily driven by mechanisms such as radial diffusion, wave-particle interactions, substorm injections, magnetospheric convection, and enhanced plasma wave activity \citep{reeves2003acceleration, tsurutani2004high, baker2014relativistic, roederer2012dynamics, thorne2021wave, tsurutani1997some}. High-speed solar wind streams (HSSs) play a critical role in energizing the magnetosphere, particularly by driving ULF wave activity, which transfers significant energy into the system and can lead to the energization of particles \citep{tsurutani1997some}. HILDCAA events are closely associated with ULF waves, making them key drivers of flux variations in the magnetosphere. ULF waves, with frequencies between 1 mHz and 1 Hz, interact with high-energy ions and electrons through drift and bounce resonances, particularly influencing particles with pitch angles near $90^\circ$. These waves affect radial diffusion processes, redistributing fluxes within the outer radiation belts and impacting relativistic electron fluxes \citep{pokhotelov2016effects}. The gradual nature of radial diffusion \citep{lejosne2020radiation} leads to delayed flux enhancements, consistent with the observed time lag between peak relativistic electron flux and solar wind speed. Additionally, particle acceleration seen in the current study shows a pitch angle dependence, with maximum energies closely tied to pitch angles. This strong pitch angle dependence underscores the critical role of ULF waves in accelerating relativistic electrons in the magnetosphere.

  \begin{table}
       \centering
       \begin{tabular}{llll}
       \hline
       \hline
    S.no. & Properties                      & ULF                                       & VLF                               \\
    \hline
    1.    & Frequency Range                 & 1 mHz to 1 Hz                             & 3 kHz to 30 kHz                    \\
    2.    & Primary Interaction Mechanisms  & Drift resonance, bounce resonance         & Cyclotron resonance, Landau damping\\
    3.    & Particle Types and Energies     & Higher energy ions and electrons          & Lower energy electrons              \\
    4.    & Pitch Angle Dependence          & More effective for particles having       & Affects a wide range of pitch       \\
          &                                 & larger pitch angles (close to 90$^\circ$) & angles, causing scattering           \\
          &                                 & and potential precipitation               &                                      \\
    \hline
    \hline
       \end{tabular}
       \caption{Summary of differences between ULF and VLF waves}
       \label{tab:Table 2}
   \end{table}
   
In contrast, VLF waves, which have frequencies between 3 kHz and 30 kHz, primarily interact with lower-energy electrons (in the keV range) through cyclotron resonance and Landau damping. Although VLF waves do not predominantly contribute to the acceleration of higher-energy particles as ULF waves do, they play a crucial role in scattering and redistributing lower-energy electrons across a range of pitch angles. This can lead to changes in the pitch angle distribution and can even result in particle precipitation into the atmosphere. Table ~\ref{tab:Table 2} summarizes the basic difference between the ULF and VLF waves for the readers for easier understanding.

Our study further underscores that during HILDCAA events, ULF waves act as primary drivers of the particle flux variation due to their ability to resonate with higher-energy particles, while VLF waves, though important, primarily affect the low-energy particle population. This aligns with findings from earlier studies like \cite{miyoshi2013high, hajra2024ultra}, which suggest that a southward interplanetary magnetic field (IMF) enhances conditions for whistler-mode chorus waves, thus accelerating electrons in the outer radiation belt. These wave-particle interactions are critical to understanding particle dynamics and flux variations during HILDCAA events, especially with ULF waves playing a dominant role.

\section{Concluding remarks}\label{sec:Conclusion}
This study investigates the variability of relativistic electron fluxes ($>1$ MeV) during HILDCAA events observed by NASA's Van Allen Probes, which provided unprecedented high-resolution particle flux measurements. The analysis shows that flux enhancements occur between 0 to 2 days after the onset of HILDCAA events, with the highest energy flux observed reaching up to $\sim$4 MeV. Notably, electrons with perpendicular pitch angles exhibited greater flux enhancement and energization compared to field-aligned electrons, indicating that ULF waves play a key role in driving radial diffusion. The observed time delay in flux enhancement, ranging from several hours to days, further supports the involvement of ULF waves in the diffusion process. Additionally, the presence of Very Low Frequency (VLF) waves during HILDCAA events \citep{hajra2015relativistic} may contribute to the acceleration of non-relativistic electrons, complementing the overall particle energization process.

\begin{acknowledgments}
The authors of the paper, acknowledge the use of data provided by NASA OmniWeb, EMFISIS, and REPT instrument onboard Van Allen Probes, WDC for Geomagnetism, Kyoto, Japan, and IMAGE. AN thanks MHRD for providing financial support and to Sardar Vallabhbhai National Institute of Technology for providing research facilities. 
\end{acknowledgments}

\vspace{5mm}

\bibliography{main}{}
\bibliographystyle{aasjournal}

\end{document}